\documentclass[fleqn,usenatbib]{mnras}
\usepackage[T1]{fontenc}
\DeclareRobustCommand{\VAN}[3]{#2}
\let\VANthebibliography\thebibliography
\def\thebibliography{\DeclareRobustCommand{\VAN}[3]{##3}\VANthebibliography}
\usepackage{graphicx}	
\usepackage{amsmath}	
\usepackage{amssymb}	
\usepackage{booktabs}   
\usepackage{pdflscape}
\usepackage{longtable}
\usepackage{tabularx}
\usepackage{blindtext}  
\usepackage{ulem}       
\usepackage{xcolor}
\colorlet{teal}{black}
\title[The evolution of the heaviest SMBHs in jetted AGNs]{The evolution of the heaviest super-massive black-holes in jetted AGNs}

\author[Diana et al.]{
Diana A.,$^{1,2}$\thanks{a.diana3@campus.unimib.it}
Caccianiga A.,$^{2}$\thanks{alessandro.caccianiga@inaf.it}
Ighina L.,$^{2,3}$
Belladitta S.,$^{2,3}$
Moretti A.,$^{2}$
Della Ceca R.$^{2}$
\\
$^{1}$Dipartimento di Fisica G. Occhialini, Università di Milano-Bicocca, Piazza della Scienza 3, 20126 Milano, Italy\\
$^{2}$INAF - Osservatorio Astronomico di Brera, via Brera 28, I-20121 Milan, Italy\\
$^{3}$DiSAT, Università degli Studi dell’Insubria, Via Valleggio 11, 22100, Como, Italy
}
\date{Accepted XXX. Received YYY; in original form ZZZ}

\pubyear{2020}
\usepackage{newtxtext,newtxmath}
\begin{document}
\label{firstpage}
\pagerange{\pageref{firstpage}--\pageref{lastpage}}
\maketitle

\begin{abstract}
We present the space density evolution, from z=1.5 up to z=5.5, of the most massive (M$\geq10^9$M$_{\sun}$) black holes hosted in jetted \textit{Active Galactic Nuclei} (AGNs). The analysis is based on a sample of 380 luminosity-selected ($\lambda$L$_{1350}\geq10^{46}$ erg s$^{-1}$ and P$_{5\text{GHz}}\geq10^{27}$ W Hz$^{-1}$) Flat Spectrum Radio Quasars (FSRQs) obtained from the \textit{Cosmic Lens All Sky Survey} (CLASS). These sources are known to be face-on jetted AGNs (i.e. blazars) and can be exploited to infer the abundance of all the (misaligned) jetted AGNs, using a geometrical argument. \textcolor{teal}{We then compare the space density of the most massive SMBHs hosted in jetted AGNs with those present in the total population (mostly composed}  by non-jetted AGNs). We find that the space density has a peak at $z\sim3$, which is \textcolor{teal}{significantly} larger than the value observed in the total AGN population with similar optical/UV luminosities ($z\sim2.2$), but not as extreme as the value previously inferred from X-ray selected blazars ($z\gtrsim4$). The jetted fraction (jetted AGNs/total AGNs) is overall consistent with the estimates in the local Universe (10--20\%) \textcolor{teal}{and at high redshift}, assuming Lorentz bulk factors $\Gamma\approx5$. \textcolor{teal}{Finally, we find a marginal decrease in the jetted fraction at high redshifts (by a factor of $\sim2$). All these evidences point toward a different evolutionary path in the jetted AGNs compared to the total AGN population.}
\end{abstract}


\section{Introduction}
\label{secInt}
Studying how Super-Massive Black Holes (SMBHs) have evolved across cosmic time is a fundamental step for our comprehension of the formation and evolution of the cosmic structures in the Universe. According to the current paradigm, the growth of the SMBH should happen during the process of accretion on a massive black hole seed present in the centre of a galaxy (e.g. \citealp{merloni_2016} for a comprehensive review). Since the accretion is also the basis of the AGN phenomenon, tracing the evolution of the AGN phase represents the most compelling way to study the SMBHs growth, \textcolor{teal}{especially at high redshifts (e.g. \citealp{volonteri_2010, valiante_2017})}. Particularly important is to establish the possible role of relativistic jets in this process \textcolor{teal}{(for a recent review on the relativistic jets in AGNs see \citealp{Blandford_2019})}. These powerful and collimated jets are \textcolor{teal}{able} to produce strong radio emission and hence the jetted AGNs are usually referred as radio-loud (RL, e.g. \citealp{best_2012}).
If the presence of a jet is the consequence of a rapidly spinning black hole (e.g. \citealp{2011MNRAS.418L..79T, Narayan2014}), we expect that RL AGNs should be very effective in transforming accreting mass into energy \citep{2011MNRAS.418L..79T, Narayan2014}. \textcolor{teal}{A pure general relativity estimate results in a radiation efficiency of $\eta\sim0.3$ for a maximal rotating BH and $\eta\sim0.1$ for a non-rotating BH (e.g. \citealp{thorne_1974})}. Therefore the black hole mass-growth in RL AGNs should be \textcolor{teal}{on average} significantly slower compared to non-jetted, or radio-quiet (RQ), AGNs. \textcolor{teal}{As a consequence, we expect the most massive SMBHs hosted in RL to be less abundant at high redshift than those hosted in the RQ counterpart, assuming that RL AGNs remain that way through all their accretion episodes. Evidence supporting this prediction, however, has not been found yet.} On the contrary, several studies have shown no clear dependence on the ratio of RL to RQ AGNs with redshift (e.g. \citealp{stern2000, Ba_ados_2015, Liu2021}) while other studies indicate that RL AGNs with the most massive SMBHs ($\geq10^{9}$~M$_{\sun}$) show a more rapid evolution, peaking earlier (z$\gtrsim$4; e.g. \citealp{Ajello_2009,2015MNRAS.446.2483S}) than those hosted in RQ AGNs (peaking at z$\sim$2, e.g. \citealp{2007ApJ...654..731H, Shen_2020}).
 
Establishing if SMBHs hosted in jetted AGNs followed a different evolutionary path compared to those hosted by non-jetted AGNs is not simple. The presence of a relativistic jet is usually inferred from the detection of a strong radio emission that reveals the radio-loud nature of an AGN (e.g. \citealp{Urry_1995}). A major problem, however, is that in RL AGNs the radio emission \textcolor{teal}{produced by the jet} is strongly dependent to the viewing angle, due to relativistic beaming, and, therefore, the same source will show a different value of radio-loudness (defined as the rest-frame radio-to-optical luminosity ratio, \citealp{kellermann1989}) depending to its actual \textcolor{teal}{orientation}. This is particularly true at high radio frequencies where the  emission is almost completely dominated by the beamed part of the radio emission (the jet).
In this respect, high-z AGNs could be very problematic since the typical observing radio frequencies (1.4--5~GHz) correspond to very high rest-frame frequencies (8--30~GHz for a z=5 AGN). 
In addition, the extended (and more isotropic) radio emission from the radio-lobes of RL AGNs could be heavily damped at high-z due to the increasing density of photons from the Cosmic Microwave Background (CMB) that interact and cool the relativistic electrons responsible for the radio emission \citep{2014MNRAS.438.2694G}. \textcolor{teal}{Finally, as in radio-quiet AGNs, obscuration of the optical/UV radiation, due to the molecular torus, makes the detection of an important fraction of sources (the so-called type-II AGNs) at high-z very challenging (e.g. \citealp{vito_2018}). }
All these observational issues can introduce relevant selection effects in all the estimates of the number of jetted AGNs at different cosmic epochs that are difficult to quantify. 

A possible way out is to focus only on the sources that are seen at small angles from the jet direction, i.e. on blazars\footnote{This category contains both sources with optical spectra characterized by strong emission lines (\textit{Flat Spectrum Radio Quasars}, FSRQs) and sources with featureless optical spectra (BL Lac objects, \citealp{Urry_1995}). Throughout this paper we will focus on the former species, therefore we will use \textit{blazar} as a synonymous for FSRQ.} (see e.g. \citealp{10.1111/j.1365-2966.2011.19024.x,2015MNRAS.446.2483S}). 
Knowing that a blazar is typically observed at a viewing angle of $\theta\lesssim1/\Gamma$, where $\Gamma$ is the Lorentz factor of the bulk velocity in the jet, we can effectively use these objects to estimate the space density of all jetted AGNs in a region of Universe, with a purely geometrical argument. Specifically, the total number of jetted AGNs (N$_{\text{jet}}$) in a given volume of Universe is expected to be: $N_{\text{jet}}\approx2\Gamma^2N_{\text{blazar}}$ i.e. there are about 200 RL AGNs for each observed blazar, assuming a typical $\Gamma=10$ (e.g. \citealp{10.1093/mnrasl/slu032}).
Another great advantage of this method is that the powerful jet is thought to wipe out the material along its path in the AGNs earlier stage, and hence obscuration effects are expected to be negligible. This makes the identification and the characterization of the optical counterpart more efficient and reliable. Therefore, from the estimate of the blazar space density it is possible to derive, in principle, a complete census of the entire jetted AGNs population without the biases due to radio emission anisotropy and to obscuration.

The application of this method requires the existence  of flux-limited and statistically complete samples of blazars.
In this paper we will use the \textit{Cosmic Lens All Sky Survey} (CLASS, \citealp{myers2003, browne2003}) to build one of the largest radio flux limited samples of blazars covering a large fraction of the sky ($\sim16300\text{ deg}^2$) and that includes sources up to z$\sim$5.5.
In particular, we are interested in tracing the evolution of the most massive SMBHs ($\geq$10$^{9}$ M$_{\sun}$) where a large difference in the evolutionary properties has been recently reported (e.g. \citealp{Ajello_2009,10.1111/j.1365-2966.2010.16449.x,10.1111/j.1365-2966.2011.19024.x,2015MNRAS.446.2483S}). 
Although for some objects in our sample the estimate of the SMBH masses is present in the literature (e.g. \citealp{Shen_2011}) all the masses are recomputed here following a single coherent method. In this way our analysis should be free from possible biases that can derive from using different techniques (or emission lines) for sources located at different redshifts. 

The paper is organized as follow. In Sect. \ref{secSample} we describe the selection process of our luminosity-limited sample. In Sect. \ref{secMass} we describe the analysis of the spectra and the estimation of the relevant parameters, furthermore we discuss the potential issues that could affect our estimates. In Sect. \ref{secDens} we derive the number density of SMBHs with M$\geq10^9$M$_{\sun}$ hosted in blazars and in all the jetted AGNs. In Sects. \ref{secDiscussion} and \ref{secConclusion} we discuss the results and we draw our conclusions.
Throughout the paper we assume a flat $\Lambda$CDM cosmology
with H$_0=70$ km s$^{-1}$Mpc$^{-1}$, $\Omega_{\lambda}=0.7$ and $\Omega_\text{M}=0.3$.

\section{The Sample}
\label{secSample}
We start our selection from the CLASS, a radio survey at 5 GHz of flat-spectrum radio sources that covers most of the northern sky ($16300\ \text{deg}^2$) and that contains $\sim11000$ sources. This catalogue was built by combining the NRAO \textit{Very Large Array Sky Survey} (NVSS), at 1.4 GHz \citep{1998AJ1151693C}, with the Green-Bank Survey (GB6) at 5 GHz \citep{articleGregory} and by selecting only the objects with a flat spectrum between 1.4 and 5 GHz ($\alpha<0.5$, with $f_\nu\propto\nu^{-\alpha}$) with a final follow-up at 8.4 GHz using the \textit{Very Large Array} (VLA) in the largest configuration (A), that granted an angular resolution of $\sim0.2\arcsec$.
The sub-arcsecond accuracy of the radio sources positions proved to be crucial to find the correct optical counterpart, in particular at faint magnitudes. Since blazars are flat-spectrum radio sources, CLASS represents the most efficient and reliable starting point to select a radio flux-limited sample of these sources suitable for statistical analyses.

We have recently carried out a specific search for blazars with redshift above 4 in the CLASS survey by efficiently pre-selecting candidates from the The Panoramic Survey Telescope and Rapid Response System (Pan-STARRS1, PS1, \citealp{2016arXiv161205560C}), using the so-called drop-out technique (see \citealp{Caccianiga_2019} for details). Nearly all the candidates with a magnitude brighter than 21 (in the \textit{r}, \textit{i} or \textit{z} filter, depending on the expected redshift of the source) have been spectroscopically confirmed, producing a list of 25 z$\geq$4 AGNs. \textcolor{teal}{The spectroscopic data are gathered from the \textit{Sloan Digital Sky Survey} (SDSS, \citealp{blanton2017}; SDSS-I/II and the Baryon Oscillation Spectroscopic Survey, BOSS; R=1500 at $\lambda=3800$\AA\ and R=2500 at $\lambda=9000$\AA) when available, or from dedicated observations using the \textit{Multi-Object Double Spectrograph} (MODS) of the \textit{Large Binocular Telescope} (LBT) with the red grating (G670L, 5000-10000\AA; R=2300 at $\lambda=7600$\AA), or with the \textit{Telescopio Nazionale Galileo} (TNG) coupled with the Device Optimized for the LOw RES-olution (DOLORES, using the LR-R/LR-B grisms; respectively, R=450 at $\lambda=4500$\AA\ and R=360 at $\lambda=7500$\AA), as detailed in \citet{Caccianiga_2019}.} The analysis of the radio spectra \citep{Caccianiga_2019} and of the X-ray data (\citealp{Ighina_2019, ighina_2021}) have then confirmed the blazar nature for 22 of these objects. If we restrict the search area to the high Galactic latitudes ($|b\arcsec|\geq20\text{ deg}$) where the identification level is close to 100\%, 3 out of 22 blazars are excluded. The resulting 19 objects constitute the high-z complete sample, hereafter C19, with a sky coverage area of 13120 deg$^2$. 

To extend this sample at lower redshifts (1$<$z$<$4) we have considered the CLASS sources falling in the sky area covered by the SDSS Data Release 14 (DR14, \citealp{2018ApJS..235...42A}) in order to benefit of the large spectroscopic database available for most of the brightest objects. Since we are interested in the most massive SMBHs, that are hosted in the most luminous AGNs, the SDSS spectroscopic data are deep enough to obtain a sample with a high spectroscopic identification level.
We have thus cross-correlated CLASS with the SDSS DR14 photometric catalogue using a 1$\arcsec$ positional tolerance to guarantee that all the counterparts are recovered (see below). We have found a counterpart for about 74\% of the CLASS sources in the SDSS area (6244 out of 8389). We have then verified that the large majority (95\%) of these counterparts have a distance from the CLASS position less than 0.3$\arcsec$ thus confirming the high completeness of the radio/optical association.
About 50\% of the counterparts have a spectroscopic identification. 
This fraction, however, dramatically increases when considering the brightest sources, leading to a high identification level of the final sample, as explained below.

In order to select sources with the same radio and optical luminosities as in the C19, we can translate the radio and optical flux limits of the high-z sample of C19 (m$_{\text{AB}} \leq 21$ and S$_{\text{5GHz}}\geq30\ \text{mJy}$) into luminosity lower limits \textcolor{teal}{at lower redshifts, assuming a starting redshift of z=4}, namely: 
\begin{equation}
\label{eqlim1}
   \lambda\text{L}_\lambda(1350\text{\AA}) \gtrsim10^{46}\ \text{erg}\ \text{s}^{-1}
\end{equation}
\begin{equation}
\label{eqlim2}
   \text{P}_{\text{5GHz}} \gtrsim10^{27}\ \text{W}\ \text{Hz}^{-1}
\end{equation}
\textcolor{teal}{We also decided to consider only sources with z>1.5. The main reason for this choice is that the inclusion of lower redshift objects in the analysis would have required to consider an additional emission line for the SMBH estimate (i.e. H$\beta$, if CIV1549\AA\ is not visible; see Section \ref{secMass} for details). The use of multiple emission lines may introduce some systematic that can affect the final results. In any case, our goal is to better sample the redshift range between 2 and 4 where the blazar space density is expected to peak and, therefore, excluding sources below z=1.5 does not affect our analysis.} We also exclude the objects with z$\geq$4, that are already included in the C19 part of the sample. \\
As mentioned above, we want to analyse a FSRQ sample, hence we exclude BL Lac objects. 
Finally, we exclude the objects with low Galactic latitude (|b$\arcsec|<20$), by analogy with the selection of C19.
The resulting 1.5$<$z$<$4 sub-sample contains 361 objects, with a sky coverage area $\sim10700\ \text{deg}^2$. 

We have then evaluated the completeness of this sample. As mentioned above, not all the CLASS sources in the SDSS area are spectroscopically identified in the literature. 
We worked out a correction to account for the missing sources.

As a first step, we split the sample in 8 bins of redshift. Then, we have translated the radio and optical flux limits of the high-z sample of C19, \textcolor{teal}{starting from redshift z=4}, at lower redshifts, using the following relations:
\begin{equation}
\label{Eq_mag}
	\text{mag}_r\text{(z)} = 20.86 + 5\text{Log} \left( \frac{\text{D}_\text{L}\text{(z)}}{\text{D}_\text{L}(4)} \right) +2.5 (\alpha_\text{O}-1)\cdot \text{Log} \left( \frac{1+\text{z}}{1+4} \right)
\end{equation}
\begin{equation}
\label{Eq_rad}
	\text{S}_{\text{5GHz}}\text{(z)} = 30 \left(\frac{\text{D}_\text{L}(4)}{\text{D}_\text{L}\text{(z)}} \right)^2 \left( \frac{1+4}{1+\text{z}} \right) ^{(\alpha_\text{R}-1)} \text{mJy}
\end{equation}
where mag$_{\text{r}}$(z) is the Galactic-extinction corrected PS1 magnitude limit in the \textit{r}-filter and S$_{\text{5GHz}}$(z) is the radio flux density limit (at 5~GHz) at redshift $z$; 20.86 is the magnitude limit used in the C19 sample (21th magnitude) corrected for the average Galactic Extinction of the C19 sample;
D$_\text{L}$(z) is the luminosity distance at redshift $z$; $\alpha_\text{O}$ and $\alpha_\text{R}$ are the optical and radio spectral indices respectively. In particular, assuming $S(\nu)\propto\nu^{-\alpha}$ we use $\alpha_\text{O}=0.44$ \citep{2001AJ....122..549V} and $\alpha_\text{R}=0$\footnote{We estimate that the scatter of the optical index ($\sim0.1$) could result in a $\sim2\%$ scatter on the number of selected blazars, while the typical uncertainties on the radio spectral index in CLASS ($\sim0.35$) could cause a $\sim9\%$ scatter.}. \textcolor{teal}{The} Eqs. \ref{Eq_mag} and \ref{Eq_rad} are equivalent to Eqs. \ref{eqlim1} and \ref{eqlim2}\textcolor{teal}{, and represent \textit{de facto} a luminosity selection}. \\
We can therefore compute the radio and optical flux limits, using the value of redshift in the centre of each bin. We then compute the identification level for each bin, as the fraction of the CLASS sources above these limits in the SDSS sky area with a spectroscopic identification in the literature. 
The inverse of this number gives the multiplicative factor ($C_{\text{id}}$) to correct for the identification level. This correction range from 1.16 (at low redshifts) to 1.45, at redshift $\sim$3.5. 

Moreover, some spectroscopically identified objects present in the literature lack a SDSS spectrum. For these objects we are not able to estimate the mass of the central BH. Nevertheless, under the assumption that all these blazars (either with spectrum or not) share the same properties, we can take into account the missing spectra with a second correction. Similarly to the $C_{\text{id}}$, we calculate this correction for each bin of redshift, specifically we compute the fraction of objects with a spectrum available from SDSS, and included in the CLASS, in each redshift bin and the correction is defined as the inverse of this number (hereafter, $C_{\text{spect}}$). The correction ranges from 1.02 to 1.15. In Tab \ref{tabIDcorr} we report the values of $C_{\text{id}}$ and $C_{\text{spect}}$ for each redshift bin. 

\textcolor{teal}{The uncertainties on these coefficients are evaluated assuming a binomial distribution $B(N,p)$, where N is the total number of sources in each bin of redshift for the considered correction, and $p$ is the inverse of the true correction. Specifically, we are interested here in the compound uncertainty obtained by multiplying the two corrections. We evaluate, through a Monte Carlo simulation, that the impact of these statistical uncertainties on the final result are at most $\sim6\%$.}

Using these corrections we can obtain a reliable estimate of the actual number of blazars expected in each redshift bin. \\
For the z$>$4 sub-sample these correction are not necessary since the identification level is almost 100\% (see C19 for details). 
\subsection{Confirming the blazar nature}
By definition, the CLASS survey should only contain sources with a flat spectrum (i.e. blazars) between 1.4 and 5~GHz. However, since the spectral index estimate is based on two frequencies only, other AGNs with more complex radio spectra may be included in the CLASS. This is the case for Gigahertz Peaked Spectrum (GPS) sources, which are thought to be young and compact AGNs and that show a peak in the radio spectrum at high frequencies (1--5~GHz, \citealp{1998PASP..110..493O}). If the peak falls in the observed 1.4--5~GHz range, the corresponding source can be misinterpreted as a FSRQ. We expect that most of the sources with a GPS spectrum are not oriented sources, i.e. blazars\footnote{This is not always true, however, as demonstrated by the existence of blazars showing a peaked radio spectrum like J0906+6930 at z=5.47 \citep{romani2004, coppejans2017, An2018,an_2020}}. In order to confirm the blazar nature of the CLASS sources, we considered their X-ray properties, which can be used as an independent and reliable proxy to assess the presence of relativistic beaming (e.g. \citealp{ghisellini2015}, \citealp{Ighina_2019}). Indeed, an X-ray emission significantly higher than that expected from a RQ AGN with the same optical/UV luminosity is a robust evidence of the presence of an extra emission coming from an oriented relativistic jet. \\

To this end, we collected X-ray data from the second \textit{Swift}-XRT point sources catalog (2SXPS, \citealp{evans2020}). We found 107 detections (35\% of the sample).  After combining the X-ray fluxes with the optical properties from SDSS, we compute the $\tilde{\alpha}_{ox}$\footnote{$\tilde{\alpha}_{ox}=-0.3026$Log(L$_{10keV}$/L$_{2500}$) \textcolor{teal}{as in \citet{Ighina_2019}}.} parameter for all these AGNs. We find that the large majority of these sources ($\sim94-96\%$, depending on the assumption on the photon index, $\Gamma$=1.5--1.8) are strong X-ray emitters in the typical range observed in blazars  ($\tilde{\alpha}<$ 1.355, see e.g. \citealp{Ighina_2019}). \\ It should be noted that the $\sim$100 CLASS sources detected in the 2SXPS catalog could not be representative of the entire sample. Indeed, the 2SXPS catalogue is based on archival pointed data and, therefore, blazars could be over-represented since they have been preferentially pointed by \textit{Swift}-XRT. To avoid this possible bias, we also considered in the computation only the sources that are serendipitously detected by \textit{Swift}-XRT, i.e. sources that were not specifically pointed. Using this sample, that  contains 24 sources, we confirm a similar abundance of blazars (i.e. $\sim92-96\%$). We note that these results are consistent with what found in the z\textgreater4 C19 sample, where about 90\% of radio selected candidates were confirmed as blazars using the X-rays data \citep{Ighina_2019}. We expect this fraction is representative of the whole sample, including the sources not yet observed in the X-rays, and, therefore, we \textcolor{teal}{apply a scale correction to the low-redshift (z<4) blazar density calculated in Section \ref{secDens} to take into account the presence of this small fraction (6\%) of non-blazars sources, in the sample.}

\begin{table}
    \centering
    \begin{tabularx}{0.46\textwidth}{>{\raggedleft\arraybackslash}X>{\raggedleft\arraybackslash}X>{\raggedleft\arraybackslash}X>{\raggedleft\arraybackslash}X>{\raggedleft\arraybackslash}X}
    \toprule
       z bin & S$_{\text{5GHz}}$ & m$_{\text{AB}}$ & C$_{\text{id}}$ & C$_{\text{spect}}$\\
    \midrule
1.50 - 1.75 & 140 & 18.88 & 1.16 & 1.14 \\
1.75 - 2.00 & 108 & 19.21 & 1.18 & 1.09 \\
2.00 - 2.25 & 86 & 19.49 & 1.22 & 1.09 \\
2.25 - 2.50 & 71 & 19.74 & 1.26 & 1.08 \\
2.50 - 2.75 & 60 & 19.96 & 1.31 & 1.02 \\
2.75 - 3.00 & 51 & 20.16 & 1.34 & 1.05 \\
3.00 - 3.50 & 42 & 20.42 & 1.39 & 1.08 \\
3.50 - 4.00 & 33 & 20.72 & 1.45 & 1.15 \\
    \bottomrule
    \end{tabularx}
    \caption{\textbf{Values of C$_{\text{id}}$ and C$_{\text{spect}}$ for different redshift bins (z<4 sub-sample).} \textbf{column 1}: redshift interval; \textbf{column 2 and 3}: respectively, lower radio-flux (mJy) limit and lower magnitude limit calculated at the central redshift of each bin (see Eqs. \ref{Eq_mag}, \ref{Eq_rad}); \textbf{column 4 and 5}: Multiplicative factors that correct for the missing identification of the sample and for the lack of SDSS spectra, respectively.}
    \label{tabIDcorr}
\end{table}

\section{Estimating the masses of the SMBH\texorpdfstring{\MakeLowercase{s}}{s}}
\label{secMass}

One of the most used and reliable method to calculate the mass of the central SMBH in type-I AGNs, is based on the virial theorem applied to the Broad Line Region (BLR). Assuming a completely virialized BLR we can compute the mass of the central BH from the BLR size and from the velocity dispersion of the orbiting clouds that form it:
\begin{equation}
    M_{\text{BH}}=f\frac{{R_{\text{BLR}} \Delta V}^2}{G}
\end{equation}
Where $\Delta V$ is a measure of the velocity dispersion of the BLR clouds, and $R_{\text{BLR}}$ is a measure of the BLR size. \textit{f} is a dimensionless factor that depends on the structure and the geometry of the BLR (e.g., \citealp{Vestergaard_2006}, hereafter VP06). \\
The Doppler broadening of the line provides the necessary information about the velocity dispersion of the ionized gas where the lines are produced. The BLR radius, instead, can be inferred from the continuum (or the line) luminosity using scaling relations of the form of  $R\propto L^{\kappa}$ (e.g., \citealp{Kaspi_2000, Grier_2019}).
This relationship can therefore be used to estimate the radius of the BLR from a single luminosity measure, without the need of a continuous monitoring of the source. For this reason, this method is called \textit{single epoch} (SE).\\
Among all the possible Broad Emission Lines (BELs), just a few of them are strong enough to be used for a reliable estimate of the BH mass: H$\beta\lambda4861$, MgII$\lambda2799$ and CIV$\lambda$1549 are the most studied and used (e.g. \citealp{Vestergaard_2006, Shen_2012, bentz2015agn}). 
With an AGN sample that covers a large range of redshift, the natural choice among the aforementioned three is the triply ionized carbon line (CIV$\lambda$1549). Even though there are some concerns about the use of this line to estimate the SMBH masses (see Subsection \ref{subIssues} for a discussion of these potential issues), we decide to use this line since it is observable from z$\sim$1.5 up to z$\sim$5.5 in an optical spectrum and, therefore, it can be consistently used for the entire sample. This is a great advantage with respect of using different emission lines (or even different methods) depending on the redshift, something that can introduce unpredictable biases in the analysis of the space density versus z. \\

There are different ways to quantify the line width, the most common one being the Full Width at Half Maximum (FWHM). However, several authors (e.g. \citealp{Vestergaard_2006, Denney_2013}) suggest that the $\sigma_l$ (line dispersion) can give more reliable results for the BH mass estimate. For instance, by comparing the CIV based masses with those derived with other independent methods, \cite{Denney_2013} has concluded that \textcolor{teal}{the line dispersion can lead to BH masses with a lower scatter (<0.3 dex) if compared to FWHM, provided that high quality spectra (S/N $\gtrsim$ 10) are used.}
We decide to compute the BH mass using both the line dispersion and the FWHM, to facilitate the comparison with the literature.
In particular, we use the two relations derived by \citet{Vestergaard_2006}:
\begin{equation}
    M_{\text{BH}}(CIV)=10^{6.66}\left(\frac{FWHM(CIV)}{10^3\ \text{km s}^{-1}}\right)^2\left(\frac{\lambda L_{\lambda}(1350\text{\AA})}{10^{44}\ \text{erg s}^{-1}}\right)^{0.53} 
    \label{eq1}
\end{equation}{}
\begin{equation}
    M_{\text{BH}}(CIV)=10^{6.73}\left(\frac{\sigma_l(CIV)}{10^3\ \text{km s}^{-1}}\right)^2\left(\frac{\lambda L_{\lambda}(1350\text{\AA})}{10^{44}\ \text{erg s}^{-1}}\right)^{0.53}
    \label{eq2}
\end{equation}
where the line dispersion and the FWHM are measured in km~s$^{-1}$ and $\lambda$L$_{\lambda}(1350\text{\AA})$ is the continuum luminosity at $1350\text{\AA}$ (source rest-frame) measured in erg~s$^{-1}$. The intrinsic scatter of these relations is 0.3--0.4~dex (see \citealp{Vestergaard_2006}) and it represents the main source of uncertainty in the mass estimate.\\

\begin{landscape}
\begin{table}
    \centering
    \begin{tabular}{rrrrrrrrrrrrrr}
    \toprule
    
name &      z &  A$_V$ & \textit{r}-mag &  S$_{\text{5GHz}}$ &       $\sigma_l$ &              FWHM & Log$\lambda$L$_{1350}$ &        LogL$_{CIV}$ & LogL$_{\text{bol}}$ &  LogM$_{\sigma}$ &     LogM$_{\text{FWHM}}$ & Log$\lambda_{\text{Edd}}$ & Log$\lambda_{\text{Edd}}$ \\

     &        &        &         &         (mJy)       &     (km s$^{-1}$)      &  (km s$^{-1}$)   & (erg s$^{-1}$)   & (erg s$^{-1}$)    & (erg s$^{-1}$)     & (M$_{\sun}$)      & (M$_{\sun}$)  & [$\sigma_l$] &   [FWHM]       \\

\midrule
GB6J001115+144608 & 4.96 & 0.242 & 19.62 &  31 & $3310\pm138$ &  $7548\pm322$ & $47.326\pm0.004$ & $45.240\pm0.012$ & $47.979\pm0.004$ & $9.53\pm0.04$ & $10.18\pm0.04$ &  $0.35\pm0.04$ & $-0.30\pm0.04$ \\
GB6J012202+030951 & 4.00 & 0.161 & 20.86 &  96 &            - & $5800\pm2000$ & $46.450\pm0.070$ &                - & $47.103\pm0.070$ &             - &  $9.52\pm0.39$ &              - & $-0.51\pm0.39$ \\
GB6J083548+182519 & 4.41 & 0.141 & 21.18 &  40 &  $2191\pm89$ &  $5032\pm211$ & $46.104\pm0.029$ & $44.497\pm0.016$ & $46.757\pm0.029$ & $8.53\pm0.04$ &  $9.18\pm0.04$ &  $0.13\pm0.04$ & $-0.52\pm0.04$ \\
GB6J083945+511206 & 4.40 & 0.175 & 19.37 &  51 & $5463\pm119$ &  $5806\pm280$ & $46.795\pm0.005$ & $45.103\pm0.007$ & $47.448\pm0.005$ & $9.69\pm0.02$ &  $9.67\pm0.04$ & $-0.34\pm0.02$ & $-0.32\pm0.04$ \\
GB6J090631+693027 & 5.47 & 0.204 & 20.54 & 100 &            - &             - &                - &                - &                - & $9.30\pm0.39$ &  $9.30\pm0.39$ &              - &              - \\
GB6J091825+063722 & 4.22 & 0.163 & 19.86 &  36 & $3031\pm109$ &  $6967\pm266$ & $46.702\pm0.009$ & $44.782\pm0.014$ & $47.355\pm0.009$ & $9.13\pm0.03$ &  $9.78\pm0.03$ &  $0.13\pm0.03$ & $-0.52\pm0.03$ \\
GB6J102107+220904 & 4.26 & 0.095 & 21.21 & 108 & $3751\pm357$ &  $3097\pm325$ & $46.007\pm0.026$ & $44.562\pm0.015$ & $46.660\pm0.026$ & $8.94\pm0.08$ &  $8.71\pm0.10$ & $-0.38\pm0.08$ & $-0.15\pm0.09$ \\
GB6J102623+254255 & 5.28 & 0.079 & 21.95 & 142 & $4183\pm488$ &  $5613\pm510$ & $46.692\pm0.011$ & $45.144\pm0.022$ & $47.345\pm0.011$ & $9.40\pm0.10$ &  $9.58\pm0.08$ & $-0.16\pm0.10$ & $-0.34\pm0.08$ \\
GB6J132512+112338 & 4.42 & 0.105 & 19.41 &  62 &  $3463\pm82$ &  $3484\pm356$ & $46.568\pm0.008$ & $45.208\pm0.011$ & $47.221\pm0.008$ & $9.17\pm0.02$ &  $9.10\pm0.10$ & $-0.05\pm0.02$ &  $0.02\pm0.10$ \\
GB6J134811+193520 & 4.40 & 0.098 & 20.73 &  38 & $5359\pm280$ &  $4838\pm247$ & $46.337\pm0.013$ & $44.609\pm0.012$ & $46.990\pm0.013$ & $9.43\pm0.05$ &  $9.27\pm0.05$ & $-0.54\pm0.05$ & $-0.38\pm0.04$ \\
GB6J141212+062408 & 4.47 & 0.110 & 20.23 &  34 & $3052\pm313$ &  $7161\pm749$ & $46.494\pm0.015$ & $44.252\pm0.020$ & $47.147\pm0.015$ & $9.02\pm0.08$ &  $9.69\pm0.09$ &  $0.03\pm0.08$ & $-0.65\pm0.09$ \\
GB6J143023+420450 & 4.71 & 0.058 & 21.00 & 337 & $3541\pm186$ &  $8129\pm450$ & $46.627\pm0.011$ & $44.602\pm0.014$ & $47.280\pm0.011$ & $9.22\pm0.05$ &  $9.87\pm0.05$ & $-0.04\pm0.05$ & $-0.69\pm0.05$ \\
GB6J151002+570256 & 4.31 & 0.078 & 20.34 & 292 & $2540\pm113$ &  $2322\pm272$ & $46.532\pm0.010$ & $44.830\pm0.014$ & $47.185\pm0.010$ & $8.88\pm0.04$ &  $8.73\pm0.11$ &  $0.20\pm0.04$ &  $0.35\pm0.11$ \\
GB6J153533+025419 & 4.39 & 0.213 & 20.59 &  53 & $3461\pm219$ &  $7935\pm528$ & $46.363\pm0.013$ & $44.363\pm0.017$ & $47.016\pm0.013$ & $9.06\pm0.05$ &  $9.71\pm0.06$ & $-0.15\pm0.06$ & $-0.80\pm0.06$ \\
GB6J162956+095959 & 5.00 & 0.283 & 21.97 &  33 & $2436\pm779$ & $5613\pm1819$ & $46.314\pm0.151$ & $44.079\pm0.114$ & $46.967\pm0.151$ & $8.73\pm0.32$ &  $9.38\pm0.33$ &  $0.14\pm0.32$ & $-0.52\pm0.32$ \\
GB6J164856+460341 & 5.36 & 0.000 & 20.31 &  36 & $2604\pm207$ & $4645\pm2100$ & $46.260\pm0.142$ & $44.601\pm0.045$ & $46.913\pm0.142$ & $8.76\pm0.11$ &  $9.19\pm0.73$ &  $0.05\pm0.10$ & $-0.38\pm0.71$ \\
GB6J171103+383016 & 4.00 & 0.221 & 20.52 &  36 & $2188\pm126$ &  $2129\pm216$ & $46.280\pm0.015$ & $44.542\pm0.014$ & $46.933\pm0.015$ & $8.62\pm0.05$ &  $8.52\pm0.10$ &  $0.21\pm0.05$ &  $0.31\pm0.10$ \\
GB6J231449+020146 & 4.11 & 0.267 & 19.59 &  84 & $3975\pm395$ &  $2903\pm267$ & $46.307\pm0.010$ & $45.164\pm0.013$ & $46.960\pm0.010$ & $9.15\pm0.08$ &  $8.81\pm0.09$ & $-0.29\pm0.08$ &  $0.05\pm0.09$ \\
GB6J235758+140205 & 4.35 & 0.191 & 20.43 &  78 & $3182\pm124$ &  $7161\pm305$ & $46.267\pm0.015$ & $44.327\pm0.017$ & $46.920\pm0.015$ & $8.94\pm0.04$ &  $9.57\pm0.04$ & $-0.12\pm0.03$ & $-0.75\pm0.04$ \\
    \bottomrule
    \vspace{1ex}
    \end{tabular}
    \caption{\textbf{Properties of the high-redshift blazars (C19).} \textbf{column 1:} source name; \textbf{column 2:} redshift; \textbf{column 3:} total extinction in the V-band; \textbf{column 4:} magnitude in the \textit{r}-filter ; \textbf{column 5:} radio flux density measured at 5GHz; \textbf{columns 6 and 7:} CIV line dispersion and FWHM; \textbf{column 8:} continuum luminosity at 1350\AA\; \textbf{column 9:} CIV$\lambda$1549 line luminosity; \textbf{column 10:} bolometric luminosity, calculated assuming a bolometric correction K$_{\text{bol}}(1350\text{\AA})=4.5$; \textbf{columns 11 and 12:} SMBH mass, estimated using the VP06 relations (see text); \textbf{columns 13 and 14:} Eddington ratios calculated using the masses derived from the line dispersion and the FWHM, respectively. The parameters of GB6J012202+030951 are inferred from the printed spectrum, whereas the mass of GB6J090631+693027 has been calculated in \citet{romani2006}. In this table we report statistical uncertainties only. Please note that the calibration uncertainties in the VP06 are indeed the main source of scatter (0.3-0.4~dex, see text) in the mass estimate.}
\label{tabz>4}
\end{table}
\end{landscape}

\begin{figure}
    \centering
    \includegraphics[width=\linewidth]{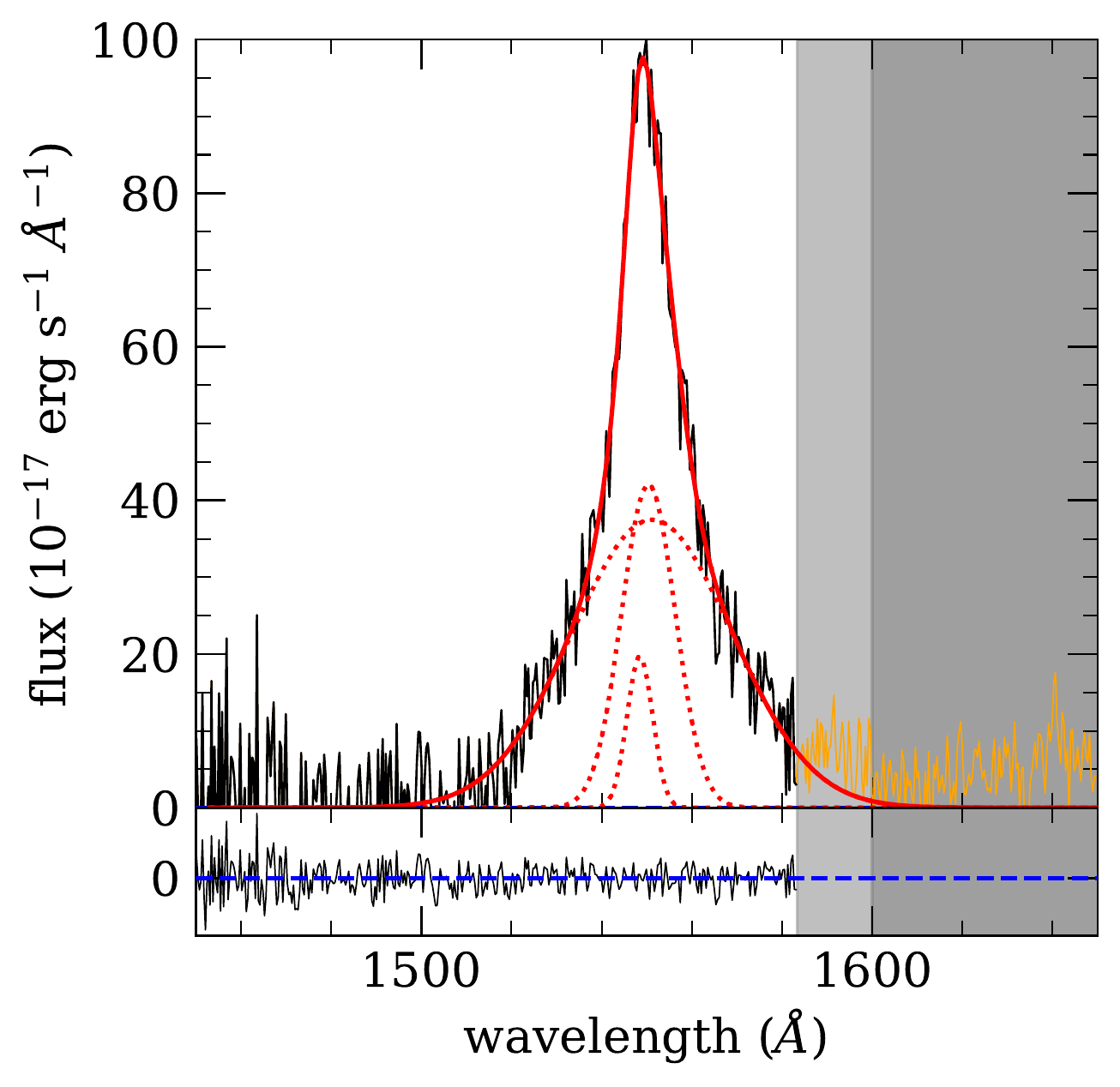}
    \caption{\textbf{Example of the fitting procedure (GB6J162030+490149).} The \textcolor{teal}{top panel} represents the rest-frame continuum-subtracted spectrum. The mask applied is represented with the grey shaded area and in orange \textcolor{teal}{is shown the excluded part of the spectrum}. The best-fitting model for the CIV line is represented with the \textcolor{teal}{solid} red line\textcolor{teal}{, while the single gaussian components are depicted with the dotted red line. The bottom panel shows the residuals.}}
    \label{figfitexample}
\end{figure}

\subsection{Spectral analysis}
\label{subsecSpectra}
The procedure used to isolate and analyze the CIV emission line follows that described in \citet{Denney_2016}.
The observed spectra are firstly corrected for Galactic extinction \citep{fitzpatrick99}.
The redshift of the source is spectroscopically determined. 
We then bring the spectrum to the source rest-frame and linearly fit the continuum around the CIV emission line, using default windows of 1435--1465\AA\ and 1690--1710\AA\ \citep{Denney_2016}. When the data within these boundaries are affected by spurious features or fall at the edge of the spectrum, other intervals are manually selected to estimate the continuum around the emission line. In many cases it was necessary to mask \textcolor{teal}{one or more} regions containing spurious features, \textcolor{teal}{such as the emission often observed between CIV and HeII$\lambda$1640 (which requires a default mask between 1600\AA\ and 1680\AA, as in \citealp{Denney_2016})}, telluric absorptions, sky line residuals and cosmic rays. 
The fitting of the CIV line is performed with a \textcolor{teal}{multi-gaussians model (1, 2 or 3 depending on the spectrum and the data quality)}. 
To better illustrate the method, we report in Fig. \ref{figfitexample} an example of the spectra preparation. \\
The fit parameters uncertainties are calculated using a Monte Carlo approach. For each object we create $N$ independent spectra based on the \textcolor{teal}{observed spectrum} with a gaussian-distributed noise added. The noise is derived from the \textit{inverse variance} vector when available, or from the RMS calculated in the continuum-subtracted spectrum around the line, when the vector was not present. For each mock spectrum the parameters are then estimated. After one thousand iterations, the standard deviation for each parameter is calculated and assumed as its uncertainty.\\ 
The best-fitting parameters with their uncertainties are reported in Table \ref{tabz>4} \footnote{\textcolor{teal}{Two} objects at z>4, i.e. GB6J012202+030951 and GB6J090631+693027, lack a SDSS spectrum, therefore for the \textcolor{teal}{former} objects we infer the parameters from the printed spectra, whereas the properties of the \textcolor{teal}{latter} object are taken from the literature \citep{romani2006}.}.
In particular, we show the best-fitting parameters of the z$>$4 sources while the values for the entire sample are reported in the Appendix.

\subsection{Eddington Ratio}
Another important parameter related to the mass of the central BH is the Eddington ratio i.e. the ratio between the bolometric luminosity and the Eddington luminosity:
\begin{equation}
    \lambda_{edd}=L_{\text{bol}}/L_{\text{Edd}}
\end{equation}
where $L_{\text{Edd}}=1.26\cdot10^{38}M/M_{\sun}$ erg s$^{-1}$ and the bolometric luminosity is the total energy produced by the AGN per unit of time integrated on all the wavelengths. 
This parameter can be estimated using a bolometric correction (K$_{\text{bol}}$) that allows the calculation of the bolometric luminosity starting from a monochromatic luminosity at a given wavelength (L$_{\text{bol}}$=K$_{\text{bol}}\cdot\lambda$L$_{\lambda}$). Here, we use an average K$_{\text{bol}}$ derived from \citet{Shen_2020}: K$_{\text{bol}}$($1450$\AA)=4.68, which can be translated at $1350$\AA~using the average spectral index, resulting in K$_{\text{bol}}$($1350$\AA)$\approx4.5$.
The resulting values of Eddington ratios of the z$>$4 objects are reported in Table \ref{tabz>4} while those of the entire sample appear in the Appendix.

\subsection{Potential issues}
\label{subIssues}
Besides the statistical errors and the intrinsic scatter of the VP06 relations, the masses derived via the SE method are potentially affected by some (possible) systematics. We briefly discuss here the most important ones trying to establish their actual relevance on the sample considered in this work. At the end of the section we will present an independent measure of the masses showing that these biases, if present, should have a modest impact on our results (also considering the large statistical uncertainties). The main potential biases that may affect our SE-derived masses are:
\begin{itemize}
    \item {\bf Orientation} - \textit{Blazars} are face-on RL AGNs. This means that we are likely observing these objects within $\sim10^{\circ}$ from the jet direction (e.g. \citealp{2010A&A...512A..24S, Ajello_2012}). It has been suggested (e.g. \citealp{Decarli_2008, Decarli2011}) that the typical BLR may have a disc-like structure and that the observed line widths are therefore significantly dependent on the particular orientation of the source since we only measure the projected component of the dispersion velocity. Therefore, in a nearly face-on source, the line width could be significantly lower than the edge-on case even if the mass of the central SMBH is the same. For a sample of randomly distributed AGNs, this effect will increase the scatter of the derived masses. In a sample of AGNs with a specific orientation, \textcolor{teal}{like blazars}, the masses could be systematically biased.\\
The SE relations (Eqs \ref{eq1} and \ref{eq2}) are calibrated on several low-redshift quasars, for which we expect a random orientation (from 0 to $\sim45$ degrees, by definition of type-I AGN). The SE relations are therefore strictly correct for a mean spatial angle, probably close to $\sim30^{\circ}$. For sources observed at lower/larger angles, the derived mass is expected to be under/over estimated, respectively. However it is not clear if all the BELs are affected by orientation. This kind of bias has been actually measured by \cite{Runnoe2014} for masses calculated using H$\beta$ but it has not been found for CIV-based masses. A similar result has been achieved by \citet{Fine2011} using a sample of RL AGNs, finding no significant correlation between the CIV line width and the BLR orientation. The proposed explanation is that the highly ionized CIV line is probably produced in a different (more isotropic) region of the BLR. Therefore, we expect that the viewing angle is not a relevant issue in our estimate. However, as discussed in the next section (see \ref{secDisk}), we have assessed this possible systematic using a completely independent method. \\
    \item {\bf \textcolor{teal}{Jet contamination and disk anisotropies}} - \textcolor{teal}{These potential issues are connected, again, with the orientation. Firstly,} the luminosity of a relativistic jet observed at small angles is significantly amplified by the relativistic beaming \citep{Urry_1995}. In principle, this emission could contaminate the continuum emitted by the accretion disk. Therefore, a continuum-luminosity based relationship, like the one we are using (Eqs.~\ref{eq1} and \ref{eq2}), may lead to a mass overestimate (e.g. \citealp{Decarli2011}). 
    \textcolor{teal}{Secondly, the continuum luminosity is known to be produced by a geometrically thin and optically thick disc-like structure, which means that the observed luminosity is higher when viewed face-on \citep{calderone_2013}. This effect could also lead to overestimate the mass, since the SE relations are calibrated on the average orientation of type-I AGNs, while we are applying them to sources observed face-on.}
    In order to evaluate the impact of these potential bias on the SE masses, we have compared the CIV line luminosities (which are not affected by \textcolor{teal}{disc-inclination effects}) against the continuum luminosities at 1350\AA\ (which could be affected by the beaming). We then compared this relation with a similar one derived by \cite{Shen_2012} from a large sample of radio-quiet AGNs \textcolor{teal}{for which the beaming is not present and that are expected to be observed, on average, at larger angles compared to blazars. If the viewing angle plays a significant role in the observed continuum luminosity}, we should observe a systematic shift of this relation with respect to the one by \cite{Shen_2012}.
    Nevertheless, comparing the luminosities of the objects in our sample with those estimated in \citet{Shen_2011} we find no evidence of a significant contamination (see Fig. \ref{figJetcontamination}). \textcolor{teal}{In particular, we can quantify the possible offset by considering the ratio $R=Log(L_{1350}/L_{CIV})$ in both the randomly aligned sample and in our sample. The two values ($R_{\text{Shen}}=1.62\pm0.28$ and $R_{\text{blazar}}=1.61\pm0.25$) are fully consistent.} \\

\begin{figure}
    \centering
    \includegraphics[width=\linewidth]{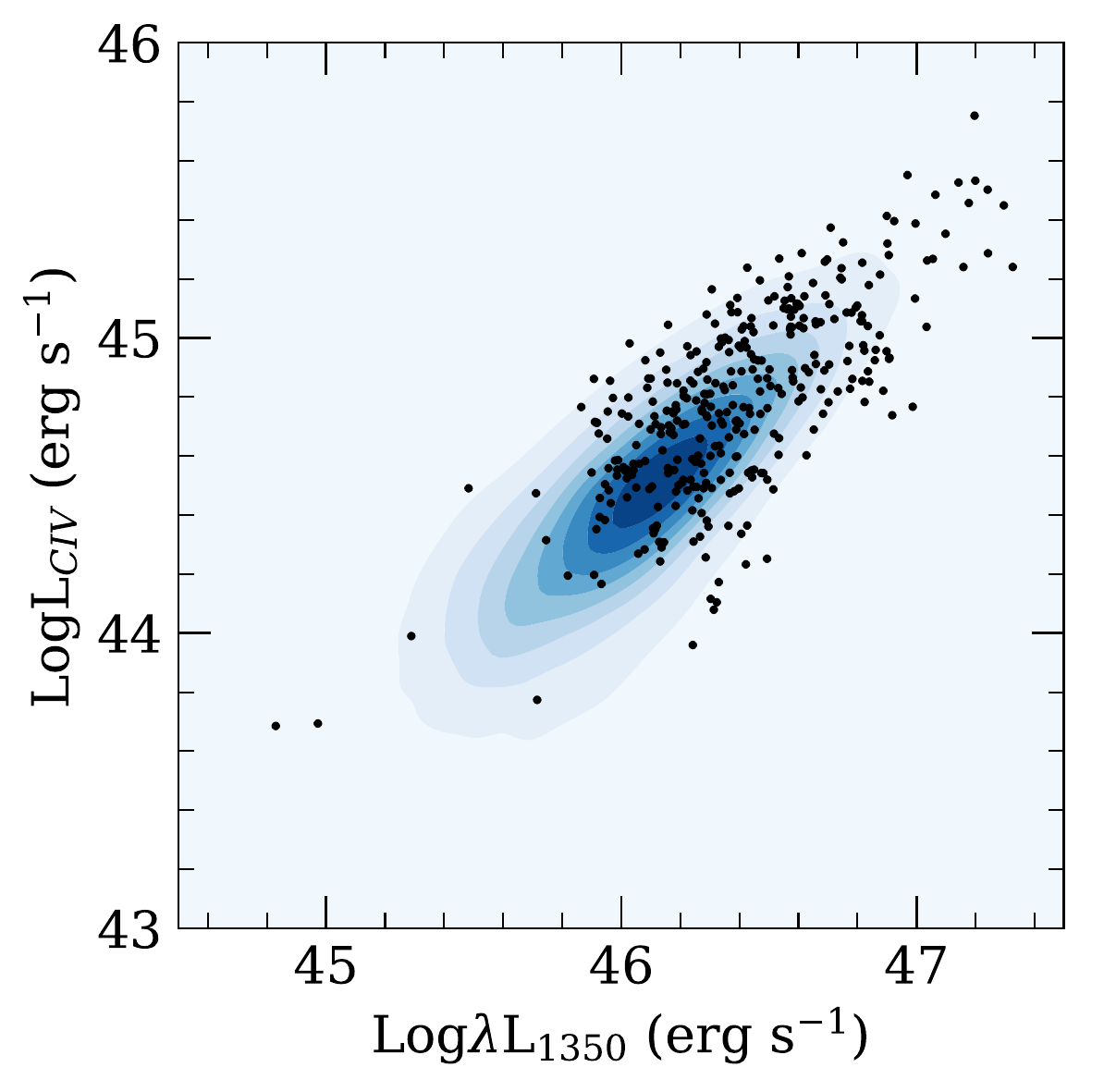}
    \caption{\textbf{Continuum luminosity versus CIV line luminosity.} The plot represents the relation between the continuum luminosities estimated at 1350\AA\ and the integrated line luminosities of the CIV, in our sample (black dots). As reference, we use the luminosities estimated in \citet{Shen_2011} using a sample of randomly oriented AGNs (represented here with the blue confidence regions, in 10\% increments). There is no evidence of any contamination from the jet, which would result in a systematic offset of the points with respect to the \citet{Shen_2011} sample.}
    \label{figJetcontamination}
\end{figure}

    \item {\bf Issues on the CIV line as a virial indicator} - The SE method is based on the assumption that the line width arises from virial motions. However, as mentioned above, CIV BELs are affected by a blueshift, although this effect seems to be less relevant in RL AGNs (\citealt{Richards2011}). Due to the high ionization energy, CIV$\lambda$1549 is likely to be produced in the innermost part of the BLR, as confirmed by reverberation mapping observations of the CIV time-lag \citep{Sun_2018}, and this effect may imply the presence of a non virialized component of the BLR. Indeed, the radiation pressure from the accretion disc may affect the regular trajectory of the gas, with a net radial radiation flow that modifies the emission line profiles \citep{Murray_1997, Leighly_2004, Denney_2012}.
There have been many efforts to improve SE mass estimates from CIV \citep{Denney_2012, Runnoe2014, Mej_a_Restrepo_2016, Coatman_2016}. Nevertheless, \citet{Denney_2013} found that the second moment of the line ($\sigma_l$) is \textcolor{teal}{only marginally affected by blueshift and that the masses derived through this quantity are less scattered with respect to the masses derived from H$\beta$ (provided that the S/N of the spectrum is high enough, namely S/N$\gtrsim$10 around 1450\AA)}. For this reason, we decided to use the $\sigma_l$ to quantify the line widths of our sample. To allow a direct comparison with the literature, however, we also compute and report in Table \ref{tabz>4} and \textcolor{teal}{in the Appendix} the values of masses obtained using the (more common) relations based on the FWHM.
\end{itemize}
\subsection{Testing the SE masses}
\label{secDisk}
Even if the potential issues described in the previous sections are not expected to have a significant impact on our analysis, we decided to verify the presence of any possible bias on the calculated masses. To this end, we use an independent technique based on the accretion disc emission (e.g. see \citealp{calderone_2013} for a detailed description of the method). This technique assumes that the optical/UV continuum emission of the AGN is produced by an optically thick, geometrically thin accretion disc (AD) that emits according to the Shakura \& Sunyaev (1973, SS73) model. With these assumptions it is possible to derive the values of M$_{\text{BH}}$ and of the accretion rate, that are free parameters of the SS73 model, simply by fitting the optical/UV data points. In spite of its simplicity, the AD method is not routinely used to derive the SMBH masses since it requires a good data coverage, in particular around the critical region where the disc emission peaks (i.e. the rest-frame UV region). For this reason, this method is typically applied to high-z objects (z$>$3--4) for which the rest-frame UV region of the spectrum is relatively well sampled by the photometric points in the visible range. However, even in high-z sources the application of this technique could be problematic since the peak of the disc emission may fall at wavelengths bluer than the Ly$\alpha\lambda1215$ where the effect of the neutral hydrogen absorption is very important,  making the photometric points not usable. This happens, in particular, in sources hosting the least massive SMBHs ($<$10$^8$ M$_{\sun}$). \\
With the main goal of testing the SE masses, we decided to apply the AD method to the high-z (z$\geq$4) objects in the CLASS sample, for which the UV spectral range is relatively well sampled.
The typical photometric coverage  available for CLASS high-z sources is limited to the few data points from PS1 not affected by the neutral hydrogen absorption and to the observed spectrum. MID-IR points from 
the Wide-field Infrared Survey Explorer (WISE,  \citealp{2010AJ....140.1868W}) cannot be used as they may be contaminated by the jet emission.  \\
As a consequence of this limited data set, the resulting masses are poorly constrained, with typical uncertainties of $\sim$0.3--0.6 dex, that are significantly larger than the statistical error on the masses derived from SE method ($\sim$0.05 dex) and comparable or even larger than the intrinsic error related to the SE method ($\sim$0.4 dex). 
In spite of these large uncertainties, the AD method can be still used to search for possible systematics in the SE masses.
\begin{figure}
    \includegraphics[width=\linewidth]{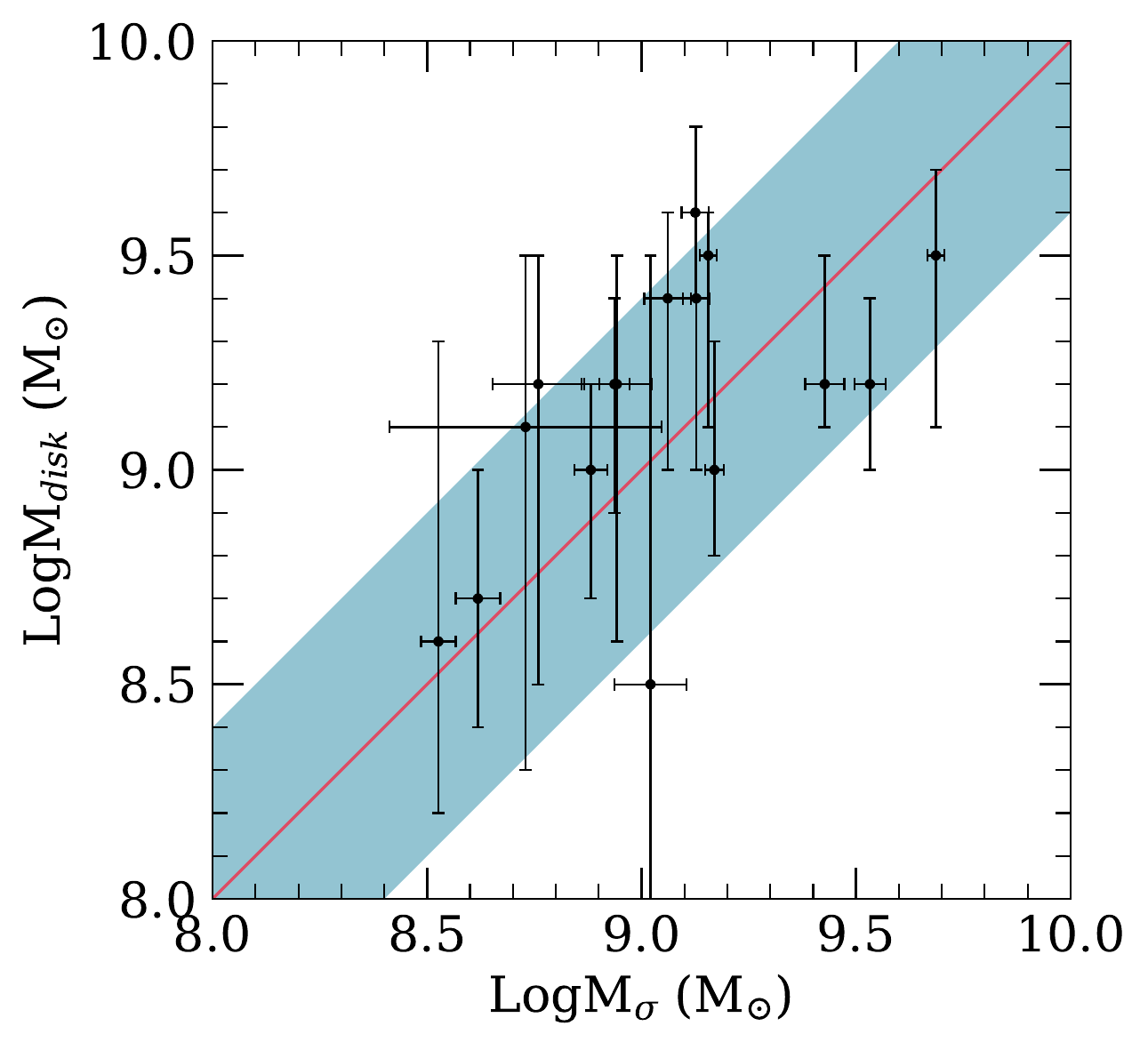}
    \caption{Comparison of the SMBH masses (z$\geq$4) derived using the SE and the AD method.
      The red solid line is the 1:1 relation; in cyan, the region corresponding to the maximum intrinsic scatter of the VP06 relations (0.4~dex). No evident bias in the mass estimate can be revealed using this independent method.} 
\label{conf_masses_sig}
\end{figure}
In Fig~\ref{conf_masses_sig} we show the comparison between the SE masses of the C19 sample (based on Eq. \ref{eq2}) and the AD masses. \\
Given the large uncertainties of the AD masses, the two values are overall in good agreement \textcolor{teal}{(Log$({M}_{\text{disk}}/{M}_{\text{SE}})=0.1\pm0.3$)}. \\
In conclusion, the masses derived with the SE method, although very uncertain, do not reveal systematic discrepancies when compared to the AD masses. This confirms that the potential biases, discussed in the previous section, if present, do not affect significantly the final results. We therefore consider the SE masses as our best estimates and we use them in the rest of the paper.

\section{Space density evolution of \texorpdfstring{$\geq10^9$M$_{\sun}$}{TEXT} SMBHs hosted by RL AGNs}
\label{secDens}

\begin{table}
    \centering
    \begin{tabular}{lccc}
        \toprule
        & Starting sample & With spectra & $M_{\text{SMBH}} \geq10^9$M$_{\sun}$ \\
        \midrule
        N sources & 380 & 352 & 243\\
        \bottomrule
    \end{tabular}
    \caption{\textbf{Number of sources selected.} The number of sources, between z=1.5 and z=5.5, in the progressively stringent sets considered. Respectively, the starting sample (\textcolor{teal}{Eqs. \ref{eqlim1}, \ref{eqlim2} and $|b\arcsec|\geq20$) composed by 361 objects with 1.5$\leq$z$<$4 plus the C19 sample (19 sources with 4$\leq$z$<$5.5)}; the objects with an available spectrum; the objects with a $M_{\text{SMBH}} \geq10^9$M$_{\sun}$ (considering the line dispersion relation, \textcolor{teal}{Eq.} \ref{eq2}).}
    \label{tablenumber}
\end{table}

An important test to assess the role of the relativistic jets in the evolution of the SMBHs across cosmic time is to compare the relative abundance of RL AGNs with respect to the total AGN population at different redshifts.
The RL fraction in the local Universe has been the subject of several publications in the last years. The commonly accepted value was assessed to be about $\sim10\%$ (e.g. \citealp{fanti1977, Condon1980, Smith1980, sramek1980, 1995ApJ...445...62H, 2002AJ....124.2364I, 10.1111/j.1365-2966.2011.19024.x}) although some more recent works have found that the RL fraction could be as high as 20--30\% (e.g. \citealp{kellermann1989,dellaceca1994,best2005, Jiang_2007, kellermann2016}). \\
However, estimating this ratio at larger redshift is not as straightforward (e.g. \citealp{stern2000,Ba_ados_2015,Yang_2016,Liu2021}). As already discussed, part of the radio emission, i.e. the one produced by the jet, is  highly anisotropic because of the relativistic beaming that strongly boosts the emission along the jet direction. This problem is particularly important at high redshifts where we typically observe the high frequency (rest-frame) emission which is usually dominated by the jet while the isotropic component (from the radio lobes) is severely attenuated due to the steepness of the spectrum. In addition, the extended emission is more difficult to detect compared to the point-like core, especially with fluxes close to the survey sensitivities. Finally, the extended emission of RL AGNs is also expected to be significantly damped at high redshift due to the increased density of the CMB photons that interact and cool the relativistic electrons in the radio lobes (see e.g. \citealp{2014MNRAS.438.2694G}). All these potential issues make it difficult to assess whether the optically selected high-z AGNs currently not detected in the radio band are truly radio-quiet sources or simply misaligned jetted AGNs. Using blazars to trace the RL AGNs population, instead, we are not directly affected by these problems and, therefore, we can provide a reliable estimate of the true fraction of RL AGNs at all redshifts.

As discussed in Section \ref{secSample}, our sample is complete for radio powers larger than P$_{5\text{GHz}}\gtrsim10^{27}$ W Hz$^{-1}$. Considering that we expect the radio luminosity in blazars to be boosted by 2 or 3 orders of magnitude \footnote{The relativistic Doppler beaming can be expressed as $\delta^{n+\alpha}$, with a Doppler factor $\delta$=1/($\Gamma$(1--$\beta$cos$\theta$)), $\alpha$ is the radio spectral index, which in our case is $\alpha_R=0$, and $n$ ranges from 2 to 3 (see e.g. \citealp{2016ApJ...827...66S}). Therefore, assuming $sin\theta=1/\Gamma$, we obtain a boost of $\Gamma^n$=10$^2$-10$^3$.}, this means that we are able to sample the population of RL AGNs down to an intrinsic radio power of P$_{5\text{GHz}}\sim10^{24}$--$10^{25}$ W Hz$^{-1}$ which is close to the typical threshold that divides radio-loud from radio-quiet AGNs (e.g. \citealp{kellermann2016, Padovani2017}). 
Considering the \textit{deboosted} radio flux, our sample is sensitive to an intrinsic radio-loudness\footnote{Defined as $R=f_{\nu}(5\text{GHz})/f_{\nu}(4400\text{\AA})$ rest-frame \citep{kellermann1989} above $\gtrsim$10, which is, again, the traditional dividing value between RL and RQ AGNs \citep{kellermann1989}}.  
This means that we should be sensitive to the bulk of the RL AGNs population up to z$\sim$5.5. 

In principle, the presence of an optical limit (mag=21 for the z$\geq$4 sample) may prevent the selection of the least (optical) luminous sources 
($\lambda$L$_{\lambda}(1350\text{\AA})<10^{46}$erg s$^{-1}$). As previously discussed, however, we are focusing on the objects hosting the largest SMBH masses (M$\geq10^9$ M$_{\sun}$, i.e. 243 sources, see Tab. \ref{tablenumber}). The limit on the optical luminosity can be translated into a limit on the bolometric luminosity:
\begin{equation}
    L_{\text{bol}} = K_{\text{bol}}(1350\text{\AA})\cdot\nu L_{\nu}(1350\text{\AA}) \simeq 5\cdot10^{46}  \text{ erg s}^{-1}
\end{equation}
For SMBH masses between 10$^9$ and 10$^{10}$ M$_{\sun}$, this limit implies that we are sensitive down to Eddington ratios of $\sim$0.1--0.2, i.e. again, to the bulk of the population of AGNs with quasar-like spectra \citep{Shen_2011}. In summary, with the CLASS blazars we are able to evaluate the true space density of RL AGNs hosting the most massive SMBHs ($M\geq10^{9}$M$_{\sun}$) in the 1.5--5.5 redshift range. 

First, for each source we calculate the comoving volume within which it could have been observed:  
\begin{equation}
    V_{\text{eff}} = \frac{A}{4\pi}V
\end{equation}
where A is the sky coverage area in steradians of the two sub-samples, i.e. 1.272$\pi$ (or $13120 \text{~deg}^2$) above z=4, 1.038$\pi$ (or $10700 \text{~deg}^2$) under z=4. $V$ is the comoving volume between $z_1$ and $z_2$. Please note that in this case it is not necessary to apply the method suggested in \citet{1980ApJ...235..694A}, since in our luminosity-limited sample each object is always detectable up to z$\sim$5.5. 
The blazar space density in each bin (z$_1$,z$_2$) is therefore calculated as:
\begin{equation}
    n_{\text{blazar}}= \frac{C_{\text{spect}} \cdot C_{\text{id}} \cdot N}{V_{\text{eff}}}
\end{equation}

\textcolor{teal}{The sources of uncertainty in the blazars space density calculation are the coefficients (C$_{\text{id}}$ and C$_{\text{spect}}$, distributed with a binomial \textit{pdf}, see Section \ref{secSample}) and the Poisson uncertainty on the number of objects above the considered mass threshold (M$\geq$10$^9$M$_{\sun}$) which, in turns, depends on the uncertainty on the SE masses (Gauss distribution with standard deviation $\sigma=0.3M_{\sun}$, see Section \ref{secMass}). In order to evaluate the impact of these sources of uncertainties we performed a Monte Carlo simulation with 10$^5$ iterations.}

\textcolor{teal}{Finally, we can consider the blazars space density inferred from the only $z\sim6$ blazar known to date ($0.0055\substack{+0.0125 \\ -0.0046}$ Gpc$^{-3}$, \citealp{belladitta2020} and Belladitta et al. submitted).}

In order to evaluate the space density of all RL AGNs hosting a SMBH with $M_{\text{BH}}\geq10^9$M$_{\sun}$ ($n_{\text{jet}}$), as mentioned in Sec. \ref{secInt}, we use a simple geometrical argument:
\begin{equation}
    n_{\text{jet}} = 2\Gamma^2 n_{\text{blazar}}
    \label{eqjet}
\end{equation}
In this paper we assume a mean $\Gamma=10$, therefore 1 blazar every $\sim$200 RLs.
The results are reported in Tab. \ref{tabDensity} and in Fig. \ref{fig2}.
\begin{table}
\centering
\begin{tabularx}{0.46\textwidth}{p{.08\textwidth}>{\raggedleft\arraybackslash}p{.03\textwidth}>{\raggedleft\arraybackslash}p{.08\textwidth}>{\raggedleft\arraybackslash}p{.06\textwidth}>{\raggedleft\arraybackslash}p{.06\textwidth}>{\raggedleft\arraybackslash}p{.06\textwidth}}
    \toprule
        z bin & N$_{\text{bl}}$ & Log$n_{\text{bl}}$ &   & Log$n_{\text{jet}}$ & \\
    \cmidrule(lr){4-6}
        & & & & $\Gamma$ &  \\
        & & & 5 & 10 & 15 \\
    \midrule
$1.50 - 1.75$ & 26 & 0.04$\substack{ +0.10 \\ -0.11 }$ & 1.74 & 2.34 & 2.69 \\
$1.75 - 2.00$ & 26 & 0.00$\substack{ +0.10 \\ -0.11 }$ & 1.70 & 2.31 & 2.66 \\
$2.00 - 2.25$ & 37 & 0.16$\substack{ +0.08 \\ -0.09 }$ & 1.86 & 2.46 & 2.82 \\
$2.25 - 2.50$ & 30 & 0.08$\substack{ +0.09 \\ -0.10 }$ & 1.78 & 2.38 & 2.73 \\
$2.50 - 2.75$ & 34 & 0.12$\substack{ +0.08 \\ -0.09 }$ & 1.82 & 2.42 & 2.78 \\
$2.75 - 3.00$ & 31 & 0.11$\substack{ +0.08 \\ -0.09 }$ & 1.81 & 2.41 & 2.76 \\
$3.00 - 3.50$ & 33 & -0.12$\substack{ +0.08 \\ -0.09 }$ & 1.57 & 2.18 & 2.53 \\
$3.50 - 4.00$ & 14 & -0.43$\substack{ +0.12 \\ -0.13 }$ & 1.27 & 1.87 & 2.22 \\
$4.00 - 4.50$ & 8 & -0.94$\substack{ +0.15 \\ -0.24 }$ & 0.76 & 1.36 & 1.72 \\
$4.50 - 5.00$ & 2 & -1.52$\substack{ +0.37 \\ -0.45 }$ & 0.18 & 0.78 & 1.14 \\
$5.00 - 5.50$ & 2 & -1.49$\substack{ +0.37 \\ -0.45 }$ & 0.21 & 0.81 & 1.16 \\
    \bottomrule
\end{tabularx}
\caption{\textbf{Space density of the most massive ($\geq10^9M_{\sun}$) blazars in our sample.} \textbf{Column 1}: redshift interval; \textbf{column 2}: the number of blazars, within the sample, with a mass $M\geq10^9$M$_{\sun}$ in each bin of redshifts in the covered sky area; \textbf{column 3}: logarithmic space density of blazars with $M\geq10^9$M$_{\sun}$ (measured in Gpc$^{-3}$); \textbf{column 4, 5 and 6}: logarithmic space density of jetted AGNs, inferred assuming three representative Lorentz bulk factors, i.e. $\Gamma= [5, 10, 15]$ (measured in Gpc$^{-3}$). The statistical uncertainties on the logarithmic jetted AGNs densities are the same as for the logarithmic blazar densities.}
\label{tabDensity}
\end{table}

\section{Discussion}
\label{secDiscussion}
So far we have derived the space density of blazars hosting the most massive SMBHs ($\geq10^9$M$_{\sun}$) at 1.5$\leq$z$\leq$5.5. 
Using a geometrical argument, assuming a viewing angle $\theta\sim 1/\Gamma$ and a typical Lorentz bulk factor $\Gamma=10$, we have inferred the space density of all the jetted AGNs powered by the most massive SMBHs, in the considered redshift interval. We want to stress that this estimate is by definition unaffected by the torus dust extinction, therefore we are including in our result both the type-I and the type-II AGNs, \textcolor{teal}{at least in the case that the obscuration is only due to a molecular torus.}
It is known that jetted AGNs represent a minority of the total population, and several works have tried to estimate the relative abundance of these objects. These estimates converge to 10--30\% in the local Universe (e.g. \citealp{fanti1977, Condon1980, Smith1980, sramek1980, 1995ApJ...445...62H, 2002AJ....124.2364I,kellermann1989,best2005, Jiang_2007, 10.1111/j.1365-2966.2011.19024.x, kellermann2016}). 
However, it is still unclear if the jetted fraction have changed with cosmic time, especially at high-z, due to the lack of consistency in the samples at different redshifts and the intrinsic difficulty of observing partially or completely obscured jetted AGNs. \\
In order to compare the most massive SMBHs hosted in jetted AGNs with those hosted by the total AGN population, we would need complete samples containing the values of SMBH masses for each source, as in our sample of blazars. Alternatively, it is possible to use the \textit{quasar luminosity function} (QLF) and integrate it above a certain \textit{effective} luminosity, considering that SMBH mass and bolometric luminosity are tightly related\footnote{\textcolor{teal}{This is a reasonable assumption, at least in our sample, since M$_{\text{BH}}$ and $\lambda$L$_{1350}$ are closely coupled.}}. In particular, we use the most recent QLF derived in \citet{Shen_2020} \footnote{There are proposed two different fit models (referred as \textit{Global Fit A} and \textit{B}), they only differ in dealing with the QLF faint-end slope. \textcolor{teal}{Here, we} choose the model \textit{Global Fit B}.}. This QLF is corrected for intrinsic absorption assuming the neutral hydrogen column distribution suggested in \citet{Ueda_2014}, and a redshift-dependent dust-to-gas ratio \citep{ma2016}. Therefore, without further corrections, we can directly compare our 
results with this QLF. \\
In particular we integrate the QLF with a minimum luminosity:
\begin{equation}
    n_{\text{AGN}}(z) = \int_{L^{\text{eff}}_{min}}^{\infty} n_{\text{AGN}}(z, L) dL
\end{equation}
Where $n_{\text{AGN}}(z)$ is the number density of AGNs hosting the most massive BHs ($\geq10^9$M$_{\sun}$); considering the
objects with M$\geq10^9$M$_{\sun}$ in our sample, we can define L$^{\text{eff}}_{min}$ as the minimum luminosity to obtain the same number of objects \textcolor{teal}{within our full sample of 380 objects, regardless of M$_{\text{BH}}$}, with L$_{\text{bol}}\geq$L$^{\text{eff}}_{min}$. We find L$^{\text{eff}}_{min}=8.4\cdot 10^{46}$ erg s$^{-1}$. Thus, we can integrate the QLF without assuming a specific Eddington ratio distribution.

\begin{figure}
    \includegraphics[width=\linewidth]{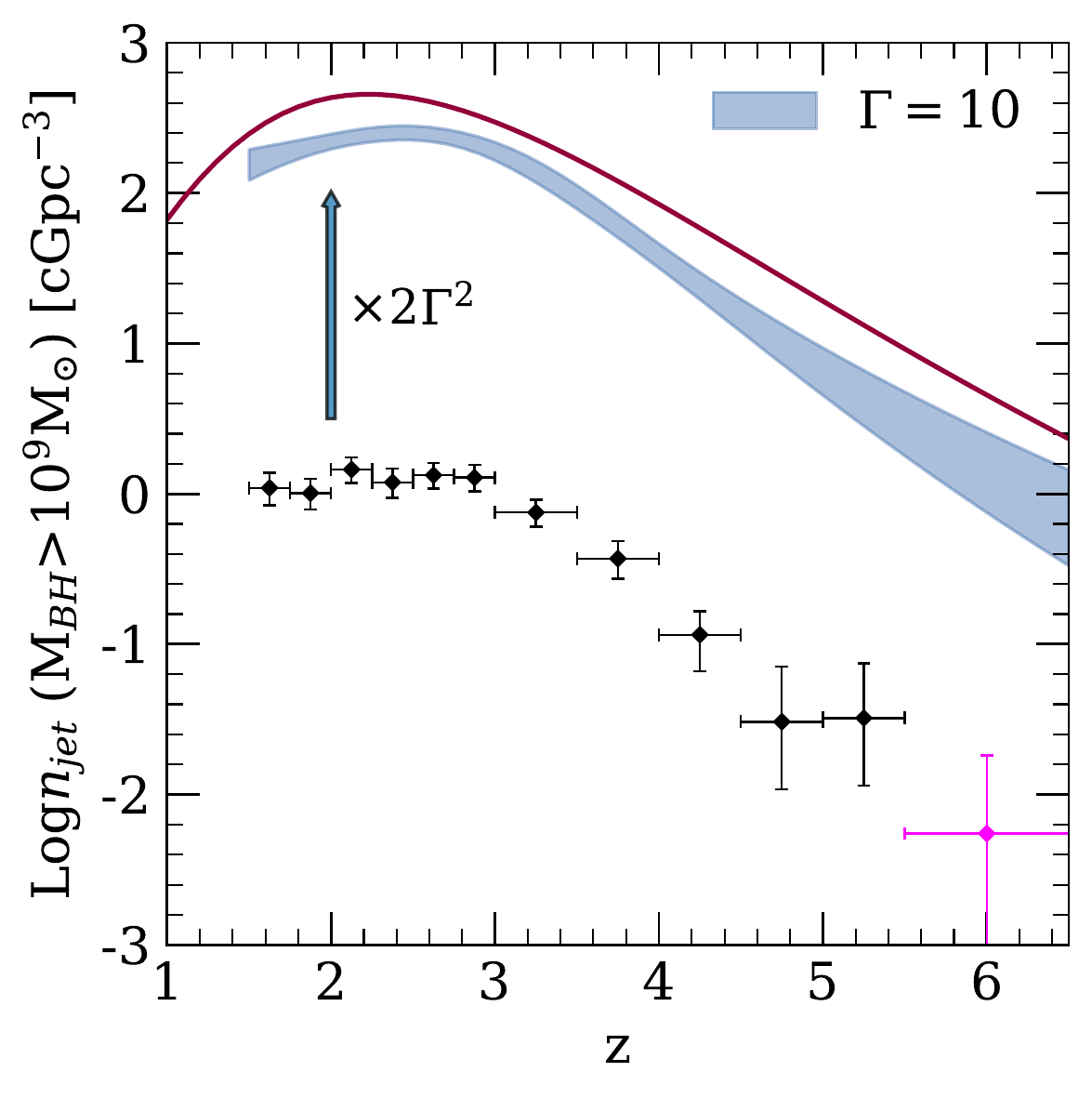}
    \caption{\textbf{Comoving space density of \textcolor{teal}{different types of AGNs} hosting black holes with $M\geq10^9$M$_{\sun}$ as a function of redshift.} \textcolor{teal}{The black dots represent the space density of SMBHs with $M\geq10^9$M$_{\sun}$ hosted by FSRQs, estimated in this work. The magenta dot represents the blazars space density inferred from the only blazar known at z>6 \citep{belladitta2020}, which is in good agreement with the extrapolation of our points. The uncertainties on the densities are estimated \textcolor{teal}{using a Monte Carlo simulation (see text)}. The blue band indicates the space density of all RL AGNs, independently of orientation (i.e. including absorbed sources), as estimated from the blazar population. It is derived by fitting the black points and then scaling up the normalization by a factor $2\Gamma^2$, assuming a mean Lorentz factor $\Gamma=10$. The delimited area represents 1$\sigma$ confidence intervals, calculated using a Monte Carlo method.
    The dark red solid line is derived from the most recent QLF \citep{Shen_2020} which estimates the space density of both jetted and non-jetted AGNs, up to z=6.5. In particular, the QLF is integrated with $L^{\text{eff}}_{\text{min}}\geq8.4\cdot 10^{46}$ erg s$^{-1}$, in order to represent only quasars similar to the bulk of our sample (see text).}}
\label{fig2}
\end{figure}

\begin{figure}
    \centering
    \includegraphics[width=\linewidth]{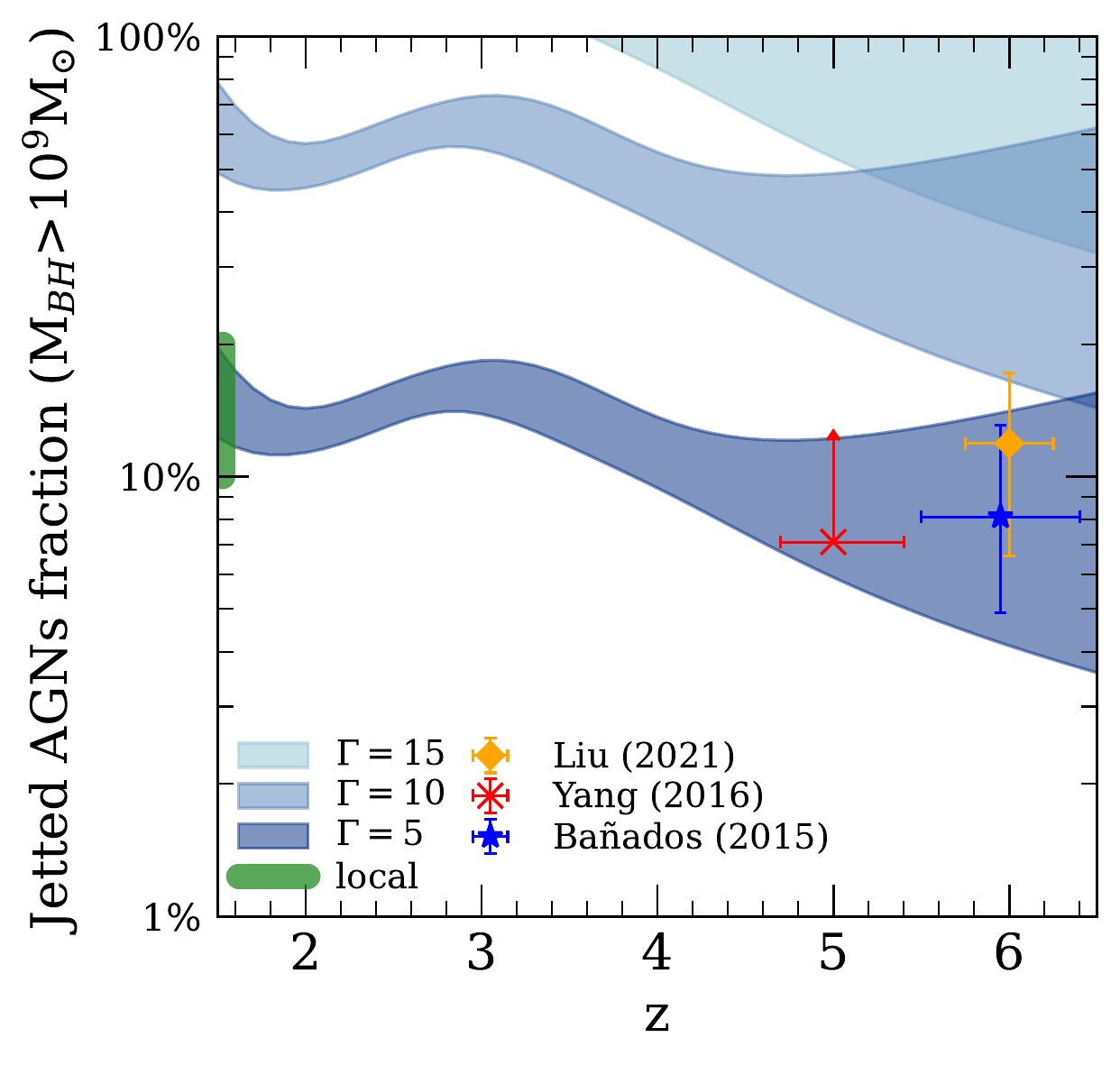}
    \caption{\textbf{Fraction of jetted AGNs hosting the most massive SMBHs ($\geq10^9$M$_{\sun}$) as a function of redshift}. We report here three different estimates assuming three values of the Lorentz factor: $\Gamma=[5, 10, 15]$. The uncertainties are estimated using a Monte Carlo method and are the same as in Fig. \ref{fig2}. Coloured dots represent previous estimates found in the literature (see the inset).}
    \label{figRL_RQ}
\end{figure}

The integrated QLF shows a luminosity dependent evolution, with a maximum density occurring at z=2.2 using the aforementioned lower luminosity (see Fig. \ref{fig2}, dark red line).  \\
The first fact that emerges from the comparison is that the space density evolution of jetted AGNs here calculated is qualitatively similar to that observed in the total AGN population. This is in contrast with previous results that found a much more rapid evolution in X-ray selected blazars with similar masses, culminating at $z\sim4$ \citep{Ajello_2009}.
\textcolor{teal}{One possible explanation for this discrepancy is that the X-ray-to-radio luminosity ratio of the jet emission strongly increases with redshift, as actually observed by several authors (e.g. \citealp{zhu_2019} and \citealp{Ighina_2019}). This increase is naturally expected if we consider the Inverse Compton interaction between the photons from the Cosmic Microwave Background (CMB) and the electrons present in the relativistic jets. This interaction is expected to produce a progressive enhancement of the X-ray emission in high-z blazars, since the CMB photon density increases as $(1+z)^4$. Using this model, \citet{ighina_2021} found that the space densities of radio-selected \citep{mao_2017} and X-ray-selected \citep{Ajello_2009} blazars can be nicely reconciled. In this picture, the apparent increase of the space density of X-ray selected blazars at high redshift is only a consequence of the progressive enhancement of their typical X-ray luminosity.} 

In spite of the similar global behaviour of the two distributions in Fig.~\ref{fig2}, we can notice some possible differences. First, the space density peak of the most massive SMBHs hosted by jetted AGNs seems to occur at slightly higher redshift (z$_{\text{peak}}\sim3$) when compared to that of the entire AGNs population (z$_{\text{peak}}\sim2.2$). In order to better quantify this difference we can fit the jetted AGNs space density \textcolor{teal}{(obtained using Eq. \ref{eqjet} with $\Gamma=10$)}, as a function of z, with a simple double power law model:
\begin{equation}
\label{fitEq}
    n(z) = \frac{2n_0}{ \left( \frac{1+z}{1+z_0} \right)^{\gamma_1} + \left( \frac{1+z}{1+z_0} \right)^{\gamma_2} }
\end{equation}
With $\gamma_1$ and $\gamma_2$ representing the two slopes, $\rho_0$ is a normalization parameter and $z_0$ represents the break point.
The uncertainties are \textcolor{teal}{calculated using a Monte Carlo simulation and are} given at 68\% confidence level. The best fit value of $z_0$ is \textcolor{teal}{$2.98\pm0.27$}. \\
\textcolor{teal}{This is not consistent with the total AGN population density peak. This in turn seems to suggest that the comoving space density of the most massive SMBHs hosted by jetted AGNs reach the maximum value about $\sim$800 Myr before those hosted by the total AGN population.}
Another possible difference between the two curves of Fig. \ref{fig2} is the flatter slope at redshifts below the peak observed in the RL AGNs compared to the total population. This is, however, an uncertain result since we are sampling only a relatively narrow range of redshifts below z<2.
\textcolor{teal}{We want to stress here that the evolutionary \textit{shape} of the jetted space density is independent of the choice of the Lorentz bulk factor, which is highly uncertain and can only change the overall normalization of the global evolutionary pattern (unless the average value of  $\Gamma$ changes across the considered range of redshifts).}

Finally, we can evaluate the evolution of the fraction of the jetted AGNs hosting the most massive SMBHs with respect to the total population. In Fig. \ref{figRL_RQ} we report the percentage fraction of the jetted AGNs estimated in our work, using three representative values of $\Gamma=[5, 10, 15]$. Clearly, given the uncertainty on the value of $\Gamma$, we cannot tightly constrain the absolute value of RL AGNs. However, assuming that the \textcolor{teal}{average} value of $\Gamma$ does not change with z, we can establish whether the fraction of jetted AGNs hosting a massive SMBH has changed between z$\sim$1 up to z$\sim$6. Fig. \ref{figRL_RQ} seems to suggest a progressive decrease of this fraction by a factor $\sim$2, going from z$\sim$1 to z$\sim$5, independently to the assumed \textcolor{teal}{average} value of $\Gamma$. This is in line with what was found by other authors (e.g. \citealp{Jiang_2007, kratzer_2015}).

\textcolor{teal}{In addition, we can compare our result with some independent estimates of the RL fraction published in the literature.} In Fig. \ref{figRL_RQ} we report the high-redshift estimates derived \textcolor{teal}{by directly counting the known RL AGNs (not necessarily blazars). We consider here only the estimates based on samples} with similar optical luminosity limit, namely: $-29< M_{1450}<-26.8$ \citep{Yang_2016}; \textcolor{teal}{PS1 $i<22.6$, at z$\sim$6} \citep{Ba_ados_2015}; $M_{1450}<-25.5$ (\citealp{Liu2021}, luminous sample) \textcolor{teal}{and with similar \textit{deboosted} radio luminosity limit $f_{\text{1.4GHz}}>1$mJy}. \textcolor{teal}{We want to stress that this method, as mentioned above, could be affected by numerous biases that can be avoided focusing on blazars instead.
\textcolor{teal}{However, considering all the uncertainties, our results are consistent with the estimates from the literature} assuming a Lorentz bulk factor between \textcolor{teal}{$\approx$6 and 8}.} \\
We can also compare our results with the observed local fraction of RL AGNs. To this end we want to consider an estimate based on a sample with similar properties, since some recent works seem to suggest that the RL fraction may depend on the mass and the luminosity of the SMBH (e.g. \citealp{cirasuolo2003, Jiang_2007, rafter2009, Yang_2016}). \citet{Jiang_2007}, in particular, gives an independent estimate of the RL fraction as a function of both luminosity and redshifts. Focusing on $z\sim1.5$ we find that the \textit{local} RL fraction is estimated to be 10\%--20\%, consistent with our low-redshift value assuming again a $\Gamma\approx5$.
\section{Conclusion}
\label{secConclusion}
We have assembled a complete and well-defined sample of 380 blazars (FSRQs) with a similar range of optical/radio luminosities ($\lambda$L$_{1350}\gtrsim10^{46}$ erg s$^{-1}$, P$_{5\text{GHz}}\gtrsim10^{27}$ W Hz$^{-1}$) hosting SMBHs with masses $\geq10^9$M$_{\sun}$, across a wide redshift interval (1.5$\leq$z$\leq$5.5).
We have used this statistically complete sample to infer the evolutionary path of all the jetted AGNs with similar properties, and we have compared the result with the total AGN population, using the most recent QLF available to date \citep{Shen_2020}. The findings can be summarized as follow:
\begin{itemize}
    \item Differently from the overall AGNs population, the jetted AGNs space density shows a peak at a slightly higher redshift ($z\sim3$ as opposed to $z\sim2.2$) but not so extreme as the value previously inferred using X-ray selected blazars ($z\sim4$).
    \item There is a marginal evidence of a flat density evolution in the jetted AGNs population at z<3 whereas the overall population density strongly decreases.
    \item The jetted AGNs fraction seems to decrease by a factor of 2 going from $z\sim1$ to $z\sim5$.
    \item Assuming a mean Lorentz bulk factor $\Gamma\approx5$ the jetted AGNs fraction is overall consistent with the values estimated in the literature in the local Universe (10--20\%) and at high redshifts.
\end{itemize}
\textcolor{teal}{The possible differences in the cosmological evolution of SMBHs hosted by jetted and non-jetted AGNs should be investigated more carefully using larger samples of blazars. In particular, in order to establish more accurately the peak position we need to improve the statistics at z$\sim$2-3. At the same time, the investigation of any difference in the slope at z>2 between the two populations requires to better sample the very high redshift end of the curve (z>5), where only few blazars have been discovered to date.
Sampling this very high-z part of the blazar population is also critical to fully understand the origin of the difference observed between radio and X-ray selected blazars. Indeed, our results seem to rule out the presence of a peak at z$\geq$4, previously suggested by X-ray selected surveys. As already explained, this discrepancy could be connected to a progressive increase of the X-ray-to-radio luminosity ratio with redshift, possibly caused by the interaction of jet with the photons from the CMB \citep{ighina_2021}. To further test this (and other) hypotheses it is fundamental to significantly increase the statistics at redshift above 5.5 where the effect of CMB is expected to be more and more relevant and where only one blazar has been discovered to date.}

\textcolor{teal}{Creating sizable samples of blazars at such high redshifts is very challenging given the particular scarcity of these sources. A big leap forward is expected in the next years thanks to the new wide area surveys, covering most of the sky, that will be carried out by the new generation of radio (\textit{Square Kilometre Array}, SKA, and its precursors), X-ray (\textit{extended Röntgen Survey with an Imaging Telescope Array}, eROSITA and Athena) and optical/IR (e.g. the \textit{Vera C. Rubin Observatory} and \textit{Euclid}) telescopes. The joint use of data covering such a large portion of the sky and of the electromagnetic spectrum has been proven to be crucial to efficiently select high-z blazars. In addition, given the extreme faintness of the expected counterparts, a fundamental role will be also played by the incoming large optical/IR telescopes, like \textit{Extremely Large Telescope}, ELT, and \textit{James Webb Space Telescope}, JWST, that will guarantee a reliable spectroscopic follow-up. We are confident that in the near future the same kind of study discussed here will be feasible up to z$\sim$7-8 and for less massive objects allowing us to better constrain the global evolution of SMBHs in jetted AGNs and to understand the actual role of the relativistic jets in the global picture.}

\section*{Acknowledgements}
The authors would like to thank the referee who provided useful and detailed comments that helped in the manuscript refinement.
This work is based on SDSS data. Funding for the Sloan Digital Sky Survey IV has been provided by the Alfred P. Sloan Foundation, the U.S. Department of Energy Office of Science, and the Participating Institutions. SDSS-IV acknowledges support and resources from the Center for High-Performance Computing at the University of Utah. The SDSS web site is www.sdss.org. We acknowledge financial contribution from the agreement ASI-INAF n. I/037/12/0 and n.2017-14-H.0 and from INAF under PRIN SKA/CTA FORECaST.

\section*{Data Availability}
The data underlying this article are available in the article and in its online supplementary material.




\bibliographystyle{mnras}
\bibliography{bibliography} 


\clearpage
\onecolumn
\appendix
\begin{landscape}
\section{Full sample data}
\label{tabz<4} 
\begin{longtable}{rrrrrrrrrrrrrr}
\caption{\textbf{Properties of the FSRQs in our sample.} \textbf{column 1:} source name; \textbf{column 2:} redshift; \textbf{column 3:} total extinction in the V-band; \textbf{column 4:} magnitude in the \textit{r}-filter ; \textbf{column 5:} radio flux density measured at 5GHz; \textbf{columns 6 and 7:} CIV line dispersion and FWHM; \textbf{column 8:} continuum luminosity at 1350\AA\; \textbf{column 9:} CIV$\lambda$1549 line luminosity; \textbf{column 10:} bolometric luminosity, calculated assuming a bolometric correction K$_{\text{bol}}(1350\text{\AA})=4.5$; \textbf{columns 11 and 12:} SMBH mass, estimated using the VP06 relations (see text); \textbf{columns 13 and 14:} Eddington ratios calculated using the masses derived from the line dispersion and the FWHM, respectively.}\\
\hline
\toprule
name &      z &  A$_V$ & \textit{r}-mag &  S$_{\text{5GHz}}$ &       $\sigma_l$ &              FWHM & Log$\lambda$L$_{1350}$ &        LogL$_{CIV}$ & LogL$_{\text{bol}}$ &  LogM$_{\sigma}$ &     LogM$_{\text{FWHM}}$ & Log$\lambda_{\text{Edd}}$ & Log$\lambda_{\text{Edd}}$ \\

     &        &        &         &         (mJy)       &     (km s$^{-1}$)      &  (km s$^{-1}$)   & (erg s$^{-1}$)   & (erg s$^{-1}$)    & (erg s$^{-1}$)     & (M$_{\sun}$)      & (M$_{\sun}$)  & [$\sigma_l$] &   [FWHM]       \\
\midrule
\endfirsthead 
\hline
\toprule
name &      z &  A$_V$ & \textit{r}-mag &  S$_{\text{5GHz}}$ &       $\sigma_l$ &              FWHM & Log$\lambda$L$_{1350}$ &        LogL$_{CIV}$ & LogL$_{\text{bol}}$ &  LogM$_{\sigma}$ &     LogM$_{\text{FWHM}}$ & Log$\lambda_{\text{Edd}}$ & Log$\lambda_{\text{Edd}}$ \\

     &        &        &         &         (mJy)       &     (km s$^{-1}$)      &  (km s$^{-1}$)   & (erg s$^{-1}$)   & (erg s$^{-1}$)    & (erg s$^{-1}$)     & (M$_{\sun}$)      & (M$_{\sun}$)  & [$\sigma_l$] &   [FWHM]       \\
\midrule
\endhead
\midrule
\multicolumn{13}{r}{{Continued on next page}} \\
\midrule
\endfoot

\bottomrule
\endlastfoot
GB6J000026+030706 & 2.35 & 0.148 & 18.74 &    91 &  $3091\pm161$ &   $5613\pm373$ & $46.515\pm0.005$ & $44.487\pm0.011$ & $47.168\pm0.005$ &  $9.04\pm0.04$ &  $9.49\pm0.06$ &  $0.02\pm0.04$ & $-0.42\pm0.06$ \\
GB6J000345+160536 & 2.31 & 0.198 & 19.49 &   109 &  $4802\pm539$ &   $4838\pm291$ & $46.383\pm0.006$ & $44.481\pm0.012$ & $47.036\pm0.006$ &  $9.36\pm0.09$ &  $9.29\pm0.05$ & $-0.42\pm0.09$ & $-0.36\pm0.05$ \\
GB6J000519+052349 & 1.90 & 0.212 & 16.23 &   303 &             - &              - &                - &                - &                - &              - &              - &              - &              - \\
GB6J000649+242232 & 1.68 & 0.276 & 18.79 &   189 & $4390\pm1472$ &   $4064\pm465$ & $46.260\pm0.005$ & $44.885\pm0.667$ & $46.913\pm0.005$ &  $9.21\pm0.23$ &  $9.08\pm0.11$ & $-0.40\pm0.23$ & $-0.26\pm0.12$ \\
GB6J000917+062549 & 2.69 & 0.292 & 19.42 &   156 &   $3064\pm95$ &   $4451\pm261$ & $46.323\pm0.005$ & $44.104\pm0.014$ & $46.977\pm0.005$ &  $8.93\pm0.03$ &  $9.19\pm0.05$ & $-0.06\pm0.03$ & $-0.31\pm0.05$ \\
GB6J000942+151357 & 2.24 & 0.252 & 19.74 &   121 &  $3415\pm508$ &   $3484\pm389$ & $45.746\pm0.017$ & $44.315\pm0.014$ & $46.399\pm0.017$ &  $8.72\pm0.09$ &  $8.67\pm0.10$ & $-0.42\pm0.09$ & $-0.37\pm0.10$ \\
GB6J001033+172437 & 1.60 & 0.196 & 16.70 &   989 &             - &              - &                - &                - &                - &              - &              - &              - &              - \\
GB6J001115+144608 & 4.96 & 0.242 & 19.62 &    31 &  $3310\pm138$ &   $7548\pm322$ & $47.326\pm0.004$ & $45.240\pm0.012$ & $47.979\pm0.004$ &  $9.53\pm0.04$ & $10.18\pm0.04$ &  $0.35\pm0.04$ & $-0.30\pm0.04$ \\
\end{longtable}

\end{landscape}
\clearpage

\twocolumn
\bsp	
\label{lastpage}
\end{document}